\documentclass[%
 preprint,
nofootinbib,
 amsmath,amssymb,
 aps,
prl,
 longbibliography,
 lengthcheck,%
]{revtex4-1}

\usepackage{graphicx}
\usepackage{dcolumn}
\usepackage{bm}
\usepackage{hyperref}
\usepackage{amsmath,bm}
\usepackage{lineno}

\def\ba{\begin{equation}\begin{array}{c}}
\def\ea{\end{array}\end{equation}}

\renewcommand{\Pi}{\hat P}

\begin{document}

\preprint{APS/123-QED}

\title{
Almost complete revivals in quantum many-body systems
}

\author{Igor Ermakov$^{1,2,3}$}
\email{ermakov1054@yandex.ru}
\author{Boris V. Fine$^{3,4}$}
\email{boris.fine@uni-leipzig.de}

\affiliation{$^1$ Skolkovo Institute of Science and Technology,
Skolkovo Innovation Center 3, Moscow  143026, Russia.}
\affiliation{$^2$ Department of Mathematical Methods for Quantum Technologies, Steklov Mathematical Institute of Russian Academy of Sciences,
8 Gubkina St., Moscow 119991, Russia.}
\affiliation{$^3$ Laboratory for the Physics of Complex Quantum Systems, Moscow Institute of Physics and Technology, Institutsky per. 9, Dolgoprudny, Moscow  region,  141700, Russia.}
\affiliation{$^4$ Institute for Theoretical Physics,
University of Leipzig, Brüderstrasse 16, 04103 Leipzig, Germany.}


\date{\today}

\begin{abstract}
Revivals of initial non-equilibrium states is an ever-present concern for the theory of dynamic thermalization in many-body quantum systems. Here we consider a nonintegrable lattice of interacting spins 1/2 and show how to construct a quantum state such that a given spin 1/2 is maximally polarized initially and then exhibits an almost complete recovery of the initial polarization at a predetermined moment of time. An experimental observation of such revivals may be utilized to benchmark quantum simulators with a measurement of only one local observable.
We further propose to utilize these revivals for a delayed disclosure of a secret.

\end{abstract}

\maketitle

An important task of the dynamic thermalization research is to determine the applicability limits of the standard assumption that a nonintegrable many-body quantum system reaches thermal equilibrium. A practical way to search for these limits is to propose examples exhibiting unconventional thermalization behavior or no thermalization at all. Such examples are often based on special quantum Hamiltonians that may exhibit proximity to integrability \cite{kinoshita2006quantum,banuls2011strong}, many-body localization \cite{nandkishore2015many}, glassy behavior \cite{rademaker2020slow,stein2013spin}, long-range interactions \cite{gong2013prethermalization,neyenhuis2017observation}, quantum scars\cite{turner2018weak,bernien2017probing,lin2019exact,iadecola2020quantum,chattopadhyay2020quantum,zhao2020quantum,lee2020exact,michailidis2020stabilizing}, and driving \cite{ji2011nonthermal,ji2018suppression}. An alternative way to induce unconventional thermalization behavior is to use special initial conditions without putting significant restrictions on a system itself \cite{goldstein2013time,dymarsky2019mechanism}. One example of the latter kind was recently mentioned by Dymarsky \cite{dymarsky2019mechanism}: namely, for an isolated many-body system, where a local observable $\hat{O}$ has equilibrium expectation value $\langle\hat{O}\rangle_\text{eq}=0$ but initially deviates from it, 
one can initiate a revival of a non-equilibrium value of $\langle \hat{O} \rangle$ at any given moment of time $\tau$ by using the initial state:
\begin{align}\label{dymstate}
|\tilde{\Psi}(0)\rangle=\frac{|\Psi(0)\rangle+|\Psi(-\tau)\rangle}{\sqrt{2}},
\end{align}
where $|\Psi(0)\rangle$ is a many-body wave function representing a ``conventional'' nonequilibrium state
such that $\langle\Psi(0)|\hat{O}|\Psi(0)\rangle$ is equal to its maximally possible value $\langle\hat{O}\rangle_\text{max}$, and $|\Psi(-\tau)\rangle$ is the state that the system should have at time $t = -\tau$ in order to arrive to $|\Psi(0)\rangle$ at $t = 0$. For the initial state $|\tilde{\Psi}(0)\rangle$, both the initial value $\langle\hat{O} (0)\rangle$ and the revived one  $\langle\hat{O} (\tau)\rangle$ are close to $\langle\hat{O}\rangle_\text{max}/2$.

In the present paper, we make a step beyond the ansatz (\ref{dymstate}) and show that, if $\hat{O}$ is a spin-1/2 operator, then one can construct a nonequilibrium state such that $\langle\hat{O} (0)\rangle = \langle\hat{O}\rangle_\text{max} $, while  $\langle\hat{O} (\tau)\rangle$ is exponentially close to $\langle\hat{O}\rangle_\text{max}$. This behavior is to be referred to below as ``almost complete revival'' (ACR). 
We argue that ACR may serve as an efficient tool for benchmarking the performance of quantum computers or quantum simulators. We further show that ACR can be used as a method for a {\it delayed disclosure of a secret}. 

Let us now construct the ACR state for a lattice of $L$ interacting spins $1/2$ described by spin operators $\{ S^{\alpha}_i \}$, where $i$ is the lattice index and $\alpha=x,y,z$ the spin projection index.   As a local observable we choose the $z$-projection of the spin on an arbitrary site labelled by index 1, i.e.  $\hat{O}=S^z_1$. We refer to the other $L-1$ spins as the ``reservoir''.  

Let us denote the bases of one-spin Hilbert spaces as $|1_i\rangle$ and $|0_i\rangle$, such that $\langle 1_i|S^z_i| 1_i\rangle=1/2$ and $\langle 0_i|S^z_i| 0_i\rangle=-1/2$. 
We define the basis $\mathcal{B}$ for the entire lattice as  $\mathcal{B}=\mathcal{B}^+\cup\mathcal{B}^-$, where
\begin{align}\label{basis05}
&\mathcal{B^+}=\{|1_1 \ 1_2  \dots 1_L\rangle, \ \dots \ , |1_1 \ 0_2 \dots 0_L\rangle\},\\
&\mathcal{B^-}=\{|0_1 \ 1_2  \dots 1_L\rangle, \ \dots \ , |0_1 \ 0_2 \dots 0_L\rangle\}.
\end{align}
represent the subspaces with the first spin being ``up'' or ``down'' respectively. Each of the two subspaces thus has dimension $N= 2^{L-1}$. For the entire basis $\mathcal{B}$ we also use the notation $\{|\varphi_n\rangle  \}$, where $n = 1, ..., N$ represents the basis $\mathcal{B^+}$ and $n = N+1, ..., 2N$ the basis $\mathcal{B^-}$.

We search for the ACR state such that it initially has the form of a tensor product $|\Phi_\text{ACR}(0)\rangle=|1_1\rangle\otimes|\Psi_\text{res}\rangle$, where  $|\Psi_\text{res}\rangle$ is   the state of the reservoir. Such a state can be parameterized as 
\begin{align}\label{psiin05}
|\Phi_\text{ACR}(0)\rangle=\sum\limits^{N}_{n=1}A_n|\varphi_n\rangle,
\end{align}
where $A_n$ are the complex amplitudes to be determined later. As follows from our indexing convention, amplitudes $A_n$ have non-zero values only for the basis states belonging to $\mathcal{B^+}$. This choice  guarantees that $\langle S_1^z (0) \rangle$ is equal to is maximum possible value $\langle S_1^z  \rangle_{\text{max}} = 1/2$.

To obtain ACR at time $\tau$, we now find such $A_n$ that $|\Phi_\text{ACR}(\tau)\rangle$ has the form:
\begin{align}\label{demand05}
|\Phi_\text{ACR}(\tau)\rangle \;&\equiv\; e^{-i{\cal H}\tau}|\Phi_\text{ACR}(0)\rangle\nonumber \\ 
&=\;\sum\limits^{N}_{n=1}C_n|\varphi_n\rangle\;\;+\;\;\delta \ |\varphi_{N+1}\rangle,
\end{align}
where $\cal H$ is the Hamiltonian of the system, while $\{C_n \}$ and $\delta$ are some complex amplitudes. 
The principal feature of the ansatz (\ref{demand05}) is that only one of $N$ basis vectors of $\mathcal{B}^-$ participates in the expansion with amplitude $\delta$, while the basis $\mathcal{B}^+$ is fully represented by the set of amplitudes $\{C_n \}$. As we show below this leads to the ACR.

Ansatz (\ref{demand05}) implies unambiguous prescription for finding $\{A_n \}$, $\{C_n \}$, and $\delta$. In order to do this, one needs first to define the matrix $u_{mn}$ of the time evolution operator $u\equiv e^{-iH\tau}$ in the the basis $\mathcal{B}$.
Then, to make sure that only state $|\varphi_{N+1}\rangle$ from the subspace $\mathcal{B}^-$ contributes to $|\Phi_\text{ACR}(\tau)\rangle$ one needs to satisfy to the following system of $N$ equations: 
\begin{align}\label{sys05}
\begin{cases}
u_{N+1,1}A_1+\cdots+u_{N+1,N}A_{N}=\delta\\
u_{N+2,1}A_1+\cdots+u_{N+2,N}A_{N}=0 \\
\cdots \\
u_{2N,1}A_1+\cdots+u_{2N,N}A_{N}=0.
\end{cases}
\end{align}
From this system one can find $N$ variables $\{ A_n \}$ as a function of yet unknown parameter $\delta$, and then find $\delta$ by normalizing $\{ A_n \}$. Finally, one can  substitute the result into the system of equations
\begin{align}\label{sys05c}
\begin{cases}
u_{1,1}A_1+\cdots+u_{1,N}A_{N}=C_1\\
u_{2,1}A_1+\cdots+u_{2,N}A_{N}=C_2 \\
\cdots \\
u_{N,1}A_1+\cdots+u_{N,N}A_{N}=C_N,
\end{cases}
\end{align}
thereby obtaining the set of amplitudes $\{ C_n \}$.

The central result of the present work is that the above prescription implies ACR, because, generically,  the values $|\delta|$ and all $|C_n|$ are of the order of $1/\sqrt{N}$,  and, as a result,
\begin{align}\label{sobv}
\langle S^z_1(\tau)\rangle=\frac{1}{2}\left(\sum\limits^{N}_{i=1}|C_i|^2-|\delta|^2\right) = 1/2 - {\cal O}(1/N) ,
\end{align}
which, in turn, means that the revived $\langle S^z_1(\tau)\rangle$ is exponentially close to $\langle S^z_1\rangle_{\text{max}}$.

While the estimate $|C_n| \sim 1/\sqrt{N}$ in the above construction is by no means surprising, the generic validity of $\delta \sim 1/\sqrt{N}$ requires a justification. Our justification is based mainly on the direct numerical solution of system (\ref{sys05})  but also it is supported by the following analytical argument.

The argument is based on the assumption that, in a generic nonintegrable system, the time evolution operator $u$ for sufficiently large times $\tau$ is similar to a random rotation in the $2N$-dimensional Hilbert space. The matrix $u_{mn}$ can then be viewed as being composed of a set of $2N$ normalized vectors $\{u_{1n}\}$, $\{u_{2n}\}$, etc., where the typical matrix element has absolute value $|u_{mn}| \sim 1/\sqrt{2N}$ and largely random phase.  The system of equations (\ref{sys05}) involves only half of the components of each vector $\{u_{N+1, n}\}$, $\{u_{N+2, n}\}$, etc. It implies that the $N$-dimensional vector $\{ A_n^* \}$ must be orthogonal to $N-1$ ``half-vectors'' $\{u_{N+2, n}\}$, ..., $\{u_{2N, n}\}$, while the value of $\delta$ is the projection of the half-vector $\{u_{N+1, n}\}$ onto the direction of $\{ A_n^* \}$. If the half-vector $\{u_{N+1, n}\}$ were uncorrelated with ``half-vectors'' $\{u_{N+2, n}\}$, ..., $\{u_{2N, n}\}$, then it should also be uncorrelated with $\{ A_n^* \}$, which means that the relative orientation of $\{u_{N+1, n}\}$ and $\{ A_n^* \}$ is random and thus the
left-hand-side of the first equation in system (\ref{sys05}) can be estimated as $A_0 u_0 \sqrt{N}$, where $A_0 \sim 1/\sqrt{N}$ and $u_0 \sim 1/\sqrt{2N}$ are the rms values of $A_n$ and $u_{mn}$ respectively. Such an estimate indeed gives $|\delta| \sim 1/\sqrt{N}$. The same kind of estimate can also be used for each line of the system (\ref{sys05c}), which would give $|C_n| \sim 1/\sqrt{N}$.

If the above assumptions were fully correct, they would imply that, once the rms values of $|C_n|$ and $|\delta|$ averaged over different not-too-small $\tau$  were exactly equal to each other. However, our numerical tests show that, even though both $C_n$ and $\delta$ are indeed of the order $1/\sqrt{N}$, there is a systematic difference between them, which is, presumably, related to subtle correlations between $u_{mn}$, for which we have no explanation.

\begin{figure}
\includegraphics[width=\columnwidth]{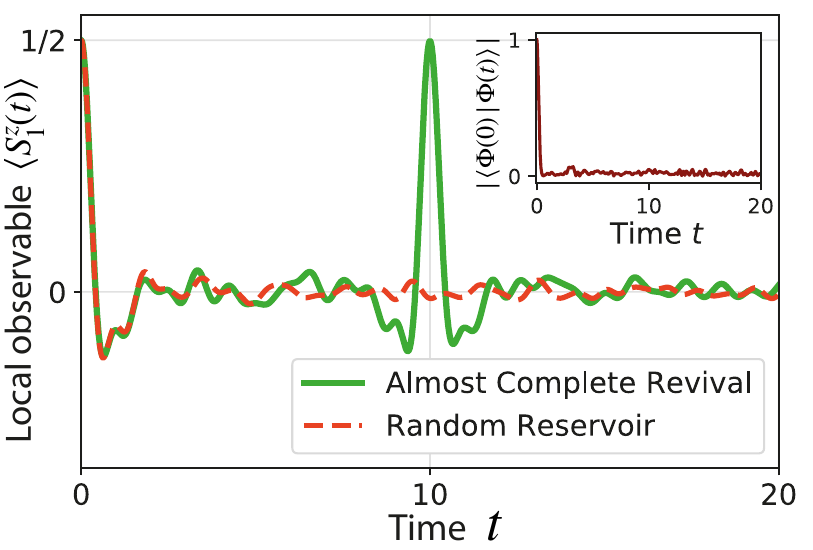}
\caption{Time evolution of the local observable $\langle S^z_1\rangle$, for a system of $L=12$ spins $1/2$ governed by the Hamiltonian (\ref{ham}). The solid green line corresponds to the ACR state devised to obtain the revival at time $\tau=10$, the dashed orange line corresponds to the random reservoir (see the text). The inset shows the fidelity $|\langle\Phi_\text{ACR}(0)|\Phi_\text{ACR}(t)\rangle|$. 
\label{s05}}
\end{figure}

\medskip
\noindent{\it Example of ACR}~

Let us now construct ACR for a translationally invariant periodic chain of $L$ spins $1/2$ described by the Hamiltonian:

\begin{align}\label{ham}
  H= & \sum_{j=1}^L\left(J_x\, S_j^x S_{j+1}^x +J_y\,S_j^y S_{j+1}^y \right)\nonumber\\
   & +\sum_{j=1}^L\left(h_x\, S_j^x+h_y\, S_j^y \right),
\end{align}
where $(J_x,J_y,h_x,h_y)=(-2.0,-4.0,2.2,2.2)$ are the interaction constants. The above values are chosen such that the Hamiltonian (\ref{ham}) is far from integrability as evidenced by the energy level-spacing statistics \cite{atas2013distribution,statistics}.

An example of the ACR behavior for the observable $S^z_1$ in a 12-spin chain is presented in Fig.~\ref{s05}. The initial state in this case was obtained by solving the system of equations (\ref{sys05}) for the revival time $\tau = 10$, and, indeed, the expected revival at $t = \tau$ was observed.

\begin{figure}
\includegraphics[width=\columnwidth]{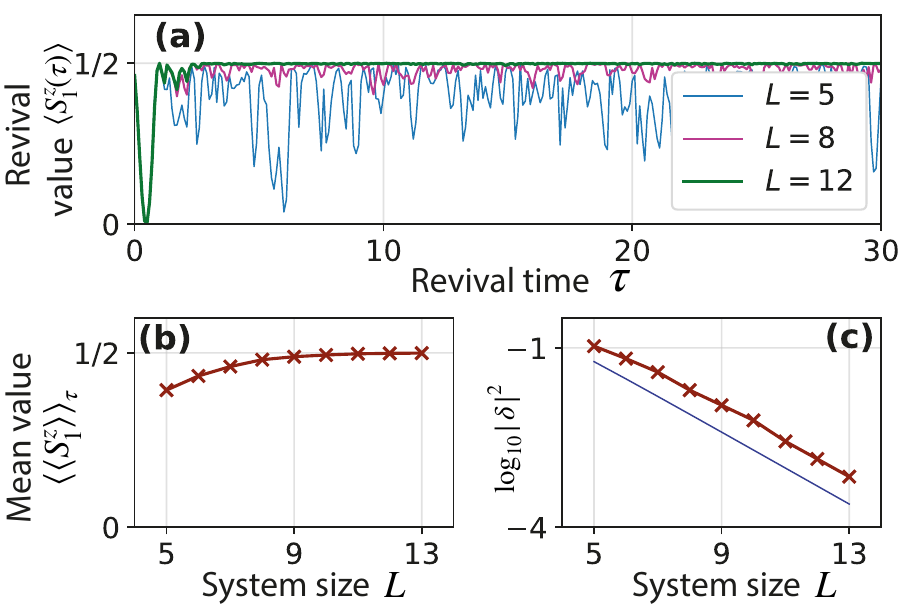}
\caption{(a) Dependence of revival value $\langle S^z_1(\tau)\rangle$ on the revival time $\tau$ for the Hamiltonian (\ref{ham}). (b)  Finite-size scaling of the $\tau$-averaged revival value $\langle\langle S^z_1(\tau)\rangle\rangle_\tau$ given by equation (\ref{s_average}). (c) Finite-size scaling of $|\delta|^2$. Crosses represent the numerical simulations, the solid blue line is the scaling $|\delta|=1/\sqrt{N}$. Averaging is performed for an interval of $\tau\in[5,30]$ with the step $\Delta t=0.01$, for $L=13$ with $\Delta t=1.0$.
\label{revfss}}
\end{figure}

In the same figure, we compare the ACR behavior with the one of a fully polarized spin in a ``random reservoir'' associated with the initial state  $|\Phi_\text{in}\rangle=|1_1\rangle\otimes|\Psi_\text{inf}\rangle$, where $|\Psi_\text{inf}\rangle$ is sampled from the infinite temperature ensemble for the remaining $L-1$ spins  \cite{gemmer2009quantum,fine2009typical}. 
As seen in Fig.~\ref{s05},  the value of $\langle S^z_1(t)\rangle$ in the case of random reservoir quickly relaxes to zero as expected for the infinite temperature equilibrium. We note that the $\langle S^z_1(t)\rangle$ for the ACR state and for the random reservoir state nearly coincide over an extended initial time interval, yet the former evolves to exhibit a revival at time $t=\tau$, while the latter shows just featureless equilibrium fluctuations. Another remarkable observation is that the almost complete revival around $t = \tau$ has the character of a nearly complete time reversal, even though the time reversal as such was not explicitly targeted by the procedure based on system (\ref{sys05}). We note in this regard that, as shown in the inset of Fig.~\ref{s05}, the fidelity of the many-body wave function $|\langle\Phi_\text{ACR} (0)|\Phi_\text{ACR}(t)\rangle|$ does not exhibit a revival at $t = \tau$. Yet, the observed time-reversed behavior of $\langle S^z_1(t)\rangle$ during ACR is consistent with the statistical argument of Ref.\cite{dykman2012large} that the most likely behavior of strong fluctuations is that of a time-reversed relaxation. 

The dependence of the revived value of $\langle S^z_1\rangle$ on the revival time $\tau$ and on the size of the lattice is illustrated in Fig.~\ref{revfss}(a): different points of the plot $\langle S^z_1(\tau)\rangle$ are obtained from different initial states $|\Phi_\text{ACR}(0)\rangle$ computed for the fixed time $\tau$ with the help of Eqs.(\ref{sys05}). There one can observe that, for smaller systems, the revived values of $\langle S^z_1\rangle$ exhibit stronger fluctuations as a function of $\tau$. However, the amplitude of these fluctuations rapidly decreases with the system size $L$. To quantify this decrease, we further observe that the fluctuation amplitudes for all system sizes $L$ are already stationary for $\tau > 5$, which allows us to characterize the typical fluctuations of $\langle S^z_1 (\tau) \rangle$ by a $\tau$-averaged  quantity $1/2 - \langle\langle S^z_1 \rangle\rangle_{\tau}$, where  
\begin{align}\label{s_average}
\langle\langle S^z_1 \rangle\rangle_{\tau}=\frac{1}{\tau_1-\tau_0}\int\limits^{\tau_1}_{\tau_0}\langle S^z_1(\tau)\rangle d\tau
\end{align}
with $\tau_0 = 5$ and $\tau_1 = 30$.
The dependence of $\langle\langle S^z_1 \rangle\rangle_{\tau}$ on the system size is plotted in Fig.~\ref{revfss}(b). Finally, in Fig.~\ref{revfss}(c), we present the semi-logarithmic plot of the $\tau$-averaged value  $\langle |\delta|^2 \rangle_{\tau} = 1/2 - \langle\langle S^z_1 \rangle\rangle_{\tau} $ as a function of $L$. This is an important plot because it shows that, for sufficiently large revival times $\tau$, the typical value of $\delta$ decreases exponentially with the system size. 

\medskip
\noindent{\it Benchmarking quantum simulators.}~

One possible application of ACR is to benchmark the performance of engineered many-qubit systems, such as quantum computers or quantum simulators.
The observation of ACR amounts to a  comprehensive test of quantum coherence and quantum control of the system.
The larger the revival time $\tau$, the more stringent is the test and the greater is the coverage of the many-qubit Hilbert space probed by the wave function in the course of the dynamical evolution. In particular, for nonintegrable systems, one can hope that the time evolution of the many-qubit wavefunction before ACR would amount to a reasonably fair sampling of the system's Hilbert space. We further note that the observation of ACR of only one qubit for sufficiently large $\tau$ indicates that the overlap between the desired initial many-qubit state and the experimentally prepared one is close to 1.

\medskip
\noindent{\it Delayed disclosure of a secret}~

Imagine that one needs to share a piece of valuable information in the form of a string of $K$ classical bits. However, the information must not be disclosed to anyone before a certain moment of time in the future. Below we propose a scheme, which allows one to implement such a delayed disclosure of a secret with the help of the ACR states.

\begin{figure}
\includegraphics[width=\columnwidth]{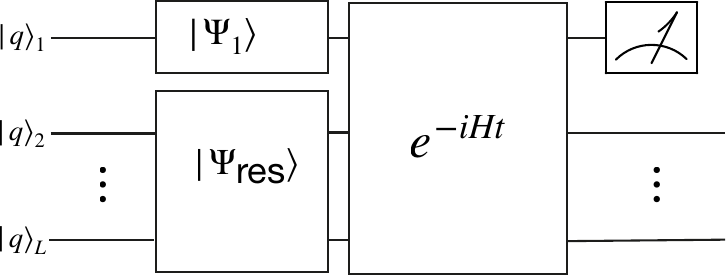}
\caption{Scheme for implementing the delayed disclosure of a secret: a classical bit is encoded into a local observable $S^z_1=\pm 1/2$. The rest of the system (a reservoir) is prepared in a state $|\Psi_\text{res}\rangle$ constructed to generate a revival at time $\tau$. The bit is retrieved when the measurement time $t$ is equal to $\tau$. Measurements at times $t$ outside of a narrow interval around $\tau$ would lead to random outcomes.
\label{qcpic}}
\end{figure}

Let us first consider only one classical bit. The state of this bit is to be encoded as $S^z_1 = \pm 1/2$ for a given spin 1/2 (a qubit), interacting with a finite ``reservoir'' of 10-50 other spins 1/2.
The state of the reservoir $|\Psi_\text{res}\rangle$ is to be prepared using the solutions of the system of equations (\ref{sys05}) such that $\langle S^z_1(t)\rangle$ exhibits a revival at $t = \tau$ (see Fig. \ref{qcpic}). After the  quantum evolution is launched, there are two possible scenarios: Either one measures $ S^z_1$ at $t=\tau$ and thereby obtains the encoded bit value with probability close to one, or the measurement is performed at a wrong time (or someone has interfered with the evolution of the system) and, therefore, the measured value of $S^z_1$ is, most likely, uncorrelated with the encoded one.

If one were to be transmitting only one bit of information, then one realization of the above procedure would not be sufficient: the occurrence of ACR on a single spin 1/2 would need to be verified either by repeating the procedure several times, or by running it simultaneously for several identical groups of spins 1/2. In this regard, the need to transmit a larger number of bits makes the verification of ACR more efficient: namely, one only needs to transmit two copies of the string of $K$ bits. If $K$ is sufficiently large and the two recovered strings are identical, then this indicates that the information was transmitted as intended.

In conclusion, we have shown how to generate an almost complete recovery of a fully polarized state of a given spin 1/2 belonging to a larger lattice of interacting spins 1/2. We have discussed possible applications of ACR to the benchmarking of quantum simulators and also proposed to utilize ACR for a delayed disclosure of a secret.

\begin{acknowledgements}
The authors are grateful to O. Lychkovskiy for useful discussions. The work of I. E. was funded by the Ministry of Science and Higher Education of the Russian Federation (grant number 075-15-2020-788)
\end{acknowledgements}

\bibliography{ref}

\end{document}